\documentclass[doublecol]{epl2} 
\usepackage{amsmath}

\title{Mapping the glassy dynamics of soft spheres onto hard-sphere behavior}
\author{Michael Schmiedeberg\inst{1,2} \and Thomas K. Haxton\inst{2}\footnote{Present address: The Molecular Foundry, Lawrence Berkeley National Laboratory, Berkeley, CA 94720, USA} \and Sidney R. Nagel\inst{3} \and Andrea J. Liu\inst{2}}
\institute{                    
\inst{1} Institut f\"ur Theoretische Physik 2: Weiche Materie, Heinrich-Heine-Universit\"at D\"usseldorf, 40225 D\"usseldorf, Germany\\
\inst{2} Department of Physics and Astronomy, University of Pennsylvania, Philadelphia, PA 19104, USA\\
\inst{3} The James Franck Institute, The University of Chicago, Chicago, IL 60637, USA
}
\shortauthor{M. Schmiedeberg \etal}

\pacs{64.70.kj}{Glasses}
\pacs{64.70.pm}{Liquids}
\pacs{64.70.pv}{Colloids}

\abstract{
We show that the dynamics of soft-sphere systems with purely repulsive interactions can be described by introducing an effective hard-sphere diameter determined using the Andersen-Weeks-Chandler approximation.  We find that this approximation, known to describe static properties of liquids, also gives a good description of a dynamical quantity, the relaxation time, even in the vicinity of the glass transition.}

\begin{document}
\maketitle

Relaxation in a liquid proceeds by numerous pathways that each depend on temperature and pressure.  It has remained a challenge to delineate different modes of relaxation and to bring them into a framework where they can be treated on a common footing~\cite{rosenfeld,dzugutov,mittal,gnan,noyola,berthier09a,berthier09b,pond,ramireznoyola}.  To reach such a goal, it is necessary to understand what combinations of variables control liquid dynamics in different regimes of temperature and pressure.  

Some insight was gained by expressing the liquid relaxation time, $\tau$, in units of a time scale, $\sqrt{m/p \sigma}$, defined by the particle mass, $m$, diameter, $\sigma$ and pressure, $p$.  We note that this is not the most obvious choice for expressing the time scale; the conventional choice is $\sqrt{m \sigma^2/\epsilon}$, where $\epsilon$ is the energy scale of the inter-particle potential.  The choice $\sqrt{m/p \sigma}$ focuses on the elementary time for a particle rearrangement event to occur driven by the pressure.  The dimensionless relaxation time can be written as a function of dimensionless variables~\cite{xu09}: 
\begin{equation}
\tau\sqrt{p \sigma/m} =  f(T/p\sigma^3, p\sigma^3/\epsilon,..),
\label{eq:taudim}
\end{equation}
where $T$ is the temperature (or thermal energy in units where Boltzmann's constant is set to unity).   Other possible dimensionless variables include the polydispersity, or quantities describing the shape of the particles or the potential.   It was shown that if we confine ourselves to monodisperse spheres with finite-ranged repulsions then there is a major simplification~\cite{xu09} when $p\sigma^3/\epsilon$ is small so that there is little particle overlap.  In that regime, the relaxation times for many systems with different interactions can be collapsed onto a single master curve that corresponds to the hard sphere limit~\cite{xu09}.  Thus,
\begin{equation}
\lim_{p \sigma^3/\epsilon \rightarrow 0} \tau \sqrt{p \sigma/m}= f_{\textnormal{HS}}(T/p\sigma^3).
\label{eq:hslimit}
\end{equation}  
In this regime, the simple interpretation can be made that relaxation in the hard-sphere limit proceeds by thermal energy working against pressure to open up free volume.

Unlike hard spheres, soft spheres can rearrange by crossing the potential barriers imposed by their neighbors.  Once soft spheres begin to overlap, the dynamics deviate from those expected in the hard-sphere limit and depend sensitively on the interaction potential, as one might expect~\cite{xu09}.  Here we analyze the situation where the particle overlap is no longer negligible so that soft particles do not behave simply as hard spheres with the same diameter, as in eq.~\ref{eq:hslimit}.  Instead, the system responds as if it were made up of particles with a smaller {\em effective} hard-sphere diameter, $\sigma_{\textnormal{eff}}$.  

It has long been known that the {\em static} properties of soft repulsive spheres can be very well approximated by hard spheres with an appropriately chosen effective diameter~\cite{rowlinson64,rowlinson64b, barker67, barker70, andersen71}.   For damped soft-sphere systems, self-consistent generalized Langevin theory has been used to generalize these static results to the dynamics~\cite{noyola}.

Here, we calculate the effective diameter using the procedure proposed by Andersen, Weeks and Chandler (AWC) \cite{andersen71} for the static properties, as Guevara-Rodriguez, et al.~\cite{noyola} have done for damped spheres.  We replace the relaxation time, $\tau$ of a liquid of soft spheres at temperature, $T$, and number density, $\rho$, with an effective relaxation time $\tau_{\textnormal{eff}}$ of a liquid of hard spheres with effective diameter, $\sigma_{\textnormal{eff}}$, and pressure, $P_{\textnormal{eff}}$.  This hard-sphere system is constrained to have the same values of $\rho$, $T$ and $m$.  
The hard-sphere equation of state relates the packing fraction, $\phi$, to $T/p\sigma^3$.  Thus, the packing fraction of the effective hard-sphere system, $\phi_{\textnormal{eff}}=\pi \rho \sigma_{\textnormal{eff}}^3/6$, corresponds to a certain value of $T/p_{\textnormal{eff}} \sigma_{\textnormal{eff}}^3$.  This specifies the relaxation time of the effective hard-sphere liquid: $\tau_{\textnormal{eff}} \sqrt{p_{\textnormal{eff} }\sigma_{\textnormal{eff}}/m} = f_{\textnormal{HS}} (T/p_{\textnormal{eff}}\sigma_{\textnormal{eff}}^3)$.  Our central finding is that $\tau \sqrt{p \sigma/m}=c\tau_{\textnormal{eff}} \sqrt{p_{\textnormal{eff}}\sigma_{\textnormal{eff}}/m}$, where $c=1$ in case of finite-range repulsions. For long-ranged interactions, where the particle diameter is not well-defined, $c$ depends on the potential but is independent of temperature and pressure. Therefore the relaxation time of the soft-sphere system is
\begin{equation}
\tau=c\tau_{\textnormal{eff}} \sqrt{p_{\textnormal{eff}} \sigma_{\textnormal{eff}}/p \sigma}
\label{tausofthard}
\end{equation}
where $p_{\textnormal{eff}}$ and $\sigma_{\textnormal{eff}}$ are the pressure and diameter of the hard-sphere system given by the AWC approximation and $p$ and $\sigma$ are the pressure and diameter of the original soft-sphere system.  We find that Eq.~\ref{tausofthard} describes the relaxation time reasonably well for many different repulsive potentials over a wide range of temperatures and densities, including near the glass transition.

\section{System and simulation methods}
We consider a three-dimensional system of monodisperse spheres that interact according to a pair interaction $V(r_{ij})$ where $r_{ij}$ is the distance between particles $i$ and $j$:
\begin{equation}
V(r_{ij})=\begin{cases} \frac{\epsilon}{\alpha}\left(1-\frac{r_{ij}}{\sigma}\right)^{\alpha} & {\rm \ \ for\ \ } r_{ij}<\sigma \\ 
			  0 & {\rm \ \ for\ \ r_{ij} \ge \sigma}
		\end{cases}
		\label{pot}
\end{equation}
with different exponents $\alpha \ge 0$.  We also consider the Weeks-Chandler-Anderson potential \cite{weeks71}
\begin{equation}
V(r_{ij})=\begin{cases} \frac{\epsilon}{72}\left[\left(\frac{\sigma}{r_{ij}}\right)^{12}-2\left(\frac{\sigma}{r_{ij}}\right)^6+1\right] & {\rm \ \ for\ \ } r_{ij}<\sigma \\ 
			  0 & {\rm \ \ for\ \ r_{ij} \ge \sigma}
		\end{cases}
\label{WCA}
\end{equation}
and power-law repulsions,
\begin{equation}
V(r_{ij})=\epsilon\left(\frac{r_{ij}}{\sigma}\right)^{-\beta}
\label{long}
\end{equation}
with $\beta=6$ or $12$.  In the simulations, the interactions in Eq.~\ref{long} were cut off at $2.5\sigma$ and a correction term corresponding to the force at the cutoff is subtracted from $V(r_{ij})$ such that the first derivative of the potential employed in the simulations is continuous.

To calculate the relaxation time, we performed molecular dynamics simulations at fixed pressure $p$ and temperature $T$ of systems of $N=1000$ particles, using periodic boundary conditions.  For the case of hard spheres, $\alpha=0$ in eq.~\ref{pot}, we used an event-based algorithm. 
We define the relaxation time $\tau$ as the time where the mean square displacement is $\left\langle r^2(\tau)\right\rangle=\sigma^2$.  We estimated the error bar in the relaxation time by averaging over 20 different starting conditions.  The standard deviation among these runs divided by 20 yields an error bar of about 1.4\% of the relaxation time for several different state points for the different potentials studied.  At each temperature and pressure, we also calculated the mean number density, $\rho$.

In order to detect any crystallization of our samples, we calculated the bond-orientational order parameter $Q_6$ \cite{steinhardt}.    For our soft-sphere systems, $Q_6$ never exceeded $0.27$ which is well below the value of a face-centered-cubic crystal ($Q_6^{\textnormal{fcc}}\approx 0.575$~\cite{steinhardt}).  The magnitude of $Q_6$ is comparable to values measured for the large spheres in bidisperse systems at similar relaxation times and pressures.   We observed no increase of $Q_6$ with simulation time.  In addition, the third-order invariant has values $\hat W_6 \approx 0.05$, very close to the values reported for Lennard-Jones supercooled liquids and considerably higher than for crystalline clusters~\cite{steinhardt}.  These results indicate that there is no appreciable crystallization in our samples.

\begin{figure}
\onefigure[width=\columnwidth]{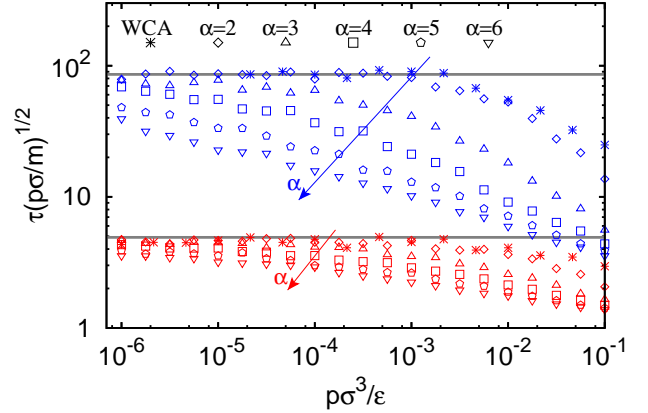}
\caption{Dimensionless relaxation time $\tau\sqrt{p\sigma/m}$, for different potentials as defined in Eqs.~\ref{pot}-{WCA}, as a function of pressure $p\sigma^3/\epsilon$ for $T/(p\sigma^3)=0.06$ (blue) and $0.2$ (red).  As $\alpha$ increases, the dimensionless relaxation time decreases, as shown by the arrows labelled $\alpha$.}
\label{fig1}
\end{figure}

\section{Analysis and Results} Figure \ref{fig1} shows the dimensionless relaxation time versus $p\sigma^3/\epsilon$ for several different potentials at two values of $T/p \sigma^3$.   For fixed $T/p \sigma^3$, the relaxation time approaches its common hard-sphere value (shown in fig.~\ref{fig1} as gray horizontal lines) at small $p \sigma^3/\epsilon$ for all potentials shown, in agreement with previous results~\cite{xu09}.  However, as $p\sigma^3/\epsilon$ increases, the relaxation time decreases in a potential-dependent fashion.  The results shown extend to pressures of order $0.1 \epsilon/\sigma^3$; for typical Lennard-Jones parameters for nonpolar molecules, where $\epsilon/k_b \sim 100-700K$ and $\sigma \sim 3-6 \AA$ \cite{data}, this corresponds to pressures on the order of several tens of atmospheres.

The first step in understanding the decrease of relaxation time in fig.~\ref{fig1} with pressure is to define an effective hard-sphere diameter to describe the liquid.  Rowlinson~\cite{rowlinson64,rowlinson64b} proposed to define $\sigma_{\textnormal{eff}}$ based on an expansion of the partition function in powers of an inverse steepness of the potential.  This approach was further developed by Barker and Henderson~\cite{barker67} and Andersen, Weeks and Chandler (AWC)~\cite{andersen71}.   The idea of the AWC approximation is to consider a set of potentials that interpolate smoothly between the effective hard-sphere potential with diameter $\sigma_{\textnormal{eff}}$ and the original potential $V(r)$.  The free energy of the original system is then expanded around the hard-sphere limit in a functional Taylor expansion.  AWC chose the effective hard-sphere diameter $\sigma_{\textnormal{eff}}$ such that the first order term in the Taylor expansion vanishes.  This yields the condition
\begin{equation}
\int d\vec r y_{\textnormal{eff}}(r) \biggl [ \exp(-V(r)/T)-\exp(-V_{\textnormal{eff}}(r)/T) \biggr ] = 0
\label{eq:ACW}
\end{equation}
where $V_{\textnormal{eff}}(r)$ is the potential for hard spheres of diameter $\sigma_{\textnormal{eff}}$ and $y_{\textnormal{eff}}(r)$ is the hard-sphere cavity pair correlation function defined by $y_{\textnormal{eff}}(r)=g_{\textnormal{eff}}(r)\exp(V_{\textnormal{eff}}(r)/T)$, where $g_{\textnormal{eff}}(r)$ is the pair correlation function of the hard-sphere system. 

We solve eq.~\ref{eq:ACW} to obtain the AWC effective hard-sphere diameter, $\sigma_{\textnormal{eff}}$. For $y_{\textnormal{eff}}(r)$, we employ the polynomial approximant introduced by Grundke and Henderson~\cite{grundke}, which uses the Verlet-Weis-approximation \cite{verlet72} for $r>\sigma_{\textnormal{eff}}$, where $y_{\textnormal{eff}}(r)=g_{\textnormal{eff}}$, and extends it to $r<\sigma_{\textnormal{eff}}$. The Grundke-Henderson approximation to $y_{\textnormal{eff}}$ has been proven to be in excellent agreement with simulation results for hard spheres\cite{torrie77}.

\begin{figure}
\onefigure[width=\columnwidth]{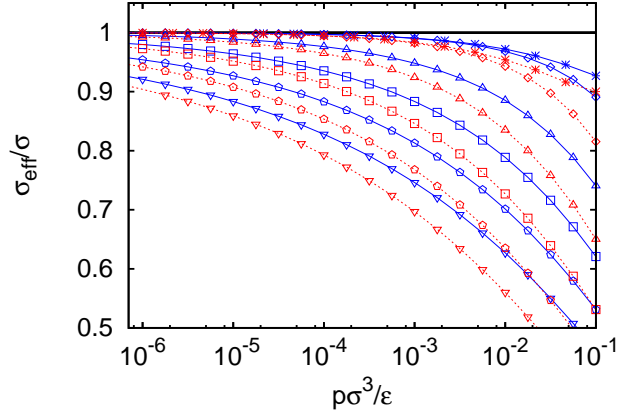}
\caption{The ratio of the effective hard-sphere diameter to the soft-sphere diameter, $\sigma_{\textnormal{eff}}/\sigma$, calculated from the Andersen-Weeks-Chandler (AWC) approximation~\cite{andersen71} as a function of $p \sigma^3/\epsilon$ at $T/p \sigma^3=0.06$ (blue, solid lines) and $T/p \sigma^3=0.2$ (red, dashed lines).   Lines are added to connect data points for the same potential at the same $T/p \sigma^3$.  Different symbols correspond to different potentials as defined in the legend of fig.~\ref{fig1}.  }
\label{fig2}
\end{figure}

The ratio of the resulting effective diameter to the original soft-sphere diameter, $\sigma_{\textnormal{eff}}/\sigma$, is plotted versus $p\sigma^3/\epsilon$ in fig.~\ref{fig2} for several different potentials.   At fixed $T/p \sigma^3$, the ratio decreases with increasing $p \sigma^3/\epsilon$.   This is expected, since the overlap increases with increasing $p \sigma^3/\epsilon$.  Additionally, fig.~\ref{fig2} shows that the effective hard-sphere diameter decreases with increasing $T/p\sigma^3$ at fixed $p \sigma^3/\epsilon$.  This is also expected, since higher overlaps, which cost more energy, are more probable at higher temperature.

While the effective hard-sphere system and the original soft-sphere system have the same temperature and number of particles, they have different pressures.  From the effective packing fraction of the hard-sphere system, $\phi_{\textnormal{eff}}$, we calculate $p_{\textnormal{eff}} \sigma_{\textnormal{eff}}^3/T$, using the Carnahan-Starling approximation~\cite{carnahan69}.  Thus, for each value of $T$ and $\rho$ studied in the original soft-sphere system, we extract a packing fraction $\phi_{\textnormal{eff}}$ and $T/p_{\textnormal{eff}} \sigma_{\textnormal{eff}}^3$ for the effective hard-sphere system.  

For finite-ranged repulsions as in Eqs.~\ref{pot}-\ref{WCA}, the potential unambiguously defines the particle diameter $\sigma$.  However, for power-law repulsions, Eq.~\ref{long}, the choice of $\sigma$ is arbitrary.  To compare our results for these different potentials, we introduce the parameter $c$. For finite-ranged repulsions we have $c=1$ while for power-law interactions we determine $c$ by a fit to Eq.~\ref{tausofthard} that  depends on the potential but does not depend on temperature or pressure. We find $c=1.31$ for $\beta=6$ and $c=1.85$ for $\beta=12$.

\begin{figure}
\onefigure[width=\columnwidth]{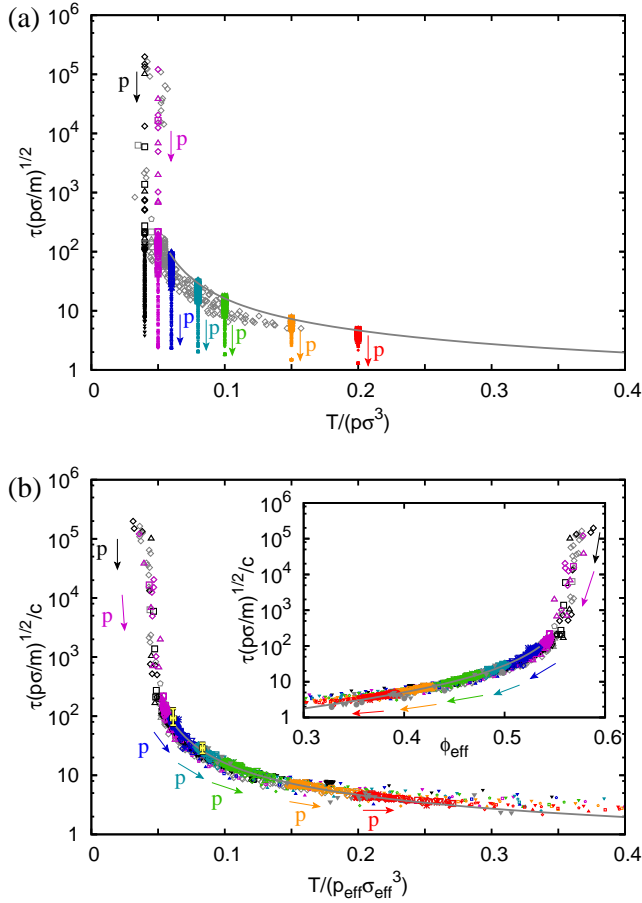}
\caption{(a) The dimensionless relaxation time $\tau\sqrt{p\sigma/m}$ versus $T/p \sigma^3$ for soft-spheres for different potentials and values of $p \sigma^3/\epsilon$. (b) Dimensionless relaxation time $\tau \sqrt{p \sigma/m}/c$ versus the effective volume fraction $\phi_{\textnormal{eff}}$ (inset) and versus $T/p_{\textnormal{eff}}\sigma_{\textnormal{eff}}^3$ (main figure).  For finite-ranged interactions the symbols for the data points are the same as in the legend to fig.~\ref{fig1}.  Different colors correspond to different values of $T/p \sigma^3$: $T/p\sigma^3=0.04$ (black), $0.05$ (purple), $0.06$ (blue), $0.08$ (cyan), $0.10$ (green), $0.15$ (orange) and $0.20$ (red).  Data points where $\sigma_{\textnormal{eff}}/\sigma>0.9$ are shown with large symbols, data with smaller $\sigma_{\textnormal{eff}}/\sigma$ are represented by small symbols.  Arrows indicate the direction of increasing pressure at each value of $T/p \sigma^3$.  Grey data points were obtained for other values of $T/p \sigma^3$ and are only shown for $\sigma_{\textnormal{eff}}/\sigma>0.9$.  Solid triangles and circles denote data with power-law interactions, using $\beta=6$ and $\beta=12$, respectively.  The typical statistical error is denoted in (b) by two yellow points.}
\label{fig3}
\end{figure}

The relaxation times for the different potentials are shown as a function of $T/p \sigma^3$ in fig.~\ref{fig3}(a). Fig.~\ref{fig3}(b) shows the dimensionless relaxation time $\tau \sqrt{p \sigma/m}/c$ versus effective packing fraction $\phi_{\textnormal{eff}}\sigma^3$ (inset) and versus $T/p_{\textnormal{eff}} \sigma_{\textnormal{eff}}^3$ (main figure). These figures show that the data collapse fairly well onto the hard-sphere curve $f_{\textnormal{HS}}$ (gray line) of eq.~\ref{eq:hslimit}. Thus, $f_{\textnormal{HS}}$ gives a reasonably good description of the relaxation time of all the systems with repulsions that we have studied.  This is a remarkable result.  It shows that for a given soft-sphere system, one can calculate an effective ratio of temperature over pressure to obtain a reasonable estimate of the dynamics, even close to the dynamical glass transition.

The collapse shown in fig.~\ref{fig3}(b) is best where the ratio $\sigma_{\textnormal{eff}}/\sigma$ is sufficiently close to unity. Large symbols in fig.~\ref{fig3} denote data where $\sigma_{\textnormal{eff}}/\sigma>0.9$ while small symbols are used for systems with higher pressure where $\sigma_{\textnormal{eff}}/\sigma\leq 0.9$.  Some of the data points with small $\sigma_{\textnormal{eff}}/\sigma$ are systematically above the hard-sphere curve (especially at high temperatures, $T/p_{\textnormal{eff}} \sigma_{\textnormal{eff}}^3>0.2$), \emph{i.e.}, the relaxation time is systematically larger than that of the corresponding hard-sphere system.  For such small $\sigma_{\textnormal{eff}}/\sigma$ values, overlaps of three or more spheres become significant, leading to corrections to the AWC approximation and a breakdown of collapse.  Note from Fig.~\ref{fig2} that all of the data for the WCA potential fall in the range where  $\sigma_{\textnormal{eff}}/\sigma> 0.9$ and are represented by large symbols; it is only for unphysically soft repulsions that there is substantial overlap at high pressures so that $\sigma_{\textnormal{eff}}/\sigma \leq 0.9$.

Because it is easier to obtain data points with high values of $\tau\sqrt{p\sigma/m}$ with simulations of soft spheres at higher pressures than for the corresponding hard-sphere system, we can use soft-sphere calculations to extend the relaxation-time curve for hard spheres.  This is shown in fig.~\ref{fig3}, where the gray curve, representing our hard-sphere simulations, ends at $\tau \sqrt{p\sigma/m} \approx 10^2$ due to the onset of crystallization but soft-sphere results extend above $\tau \sqrt{p\sigma/m} = 10^5$.  Note, however, that the extension of the master curve $f_{\textnormal{HS}}$ to higher relaxation times relies on knowledge of the metastable liquid branch of the hard-sphere equation of state. In particular, we used the Carnahan-Starling approximation to obtain fig.~\ref{fig3}(b); this approximation must fail at sufficiently high pressures.

\section{Discussion}
We note that the scaling collapse shown in Fig.~\ref{fig3} is different from the one found by Berthier and Witten~\cite{berthier09a,berthier09b}, who studied bidisperse mixtures of soft spheres.  In their analysis, the data are collapsed onto two scaling curves, one for $\phi<\phi_0$ and the other for $\phi>\phi_0$, where $\phi_0$ and the two scaling exponents are chosen to obtain the best collapse.  By contrast, the scaling collapse based on the AWC approximation collapses all of the data onto a single curve, rather than two curves.  The data shown in fig.~\ref{fig3} span a much broader range of packing fractions and temperatures than the data included by Berthier and Witten.  In addition, we have included data for many different potentials, whereas Berthier and Witten's analysis was only for the case $\alpha=2$ in eq.~\ref{pot}.  For finite-ranged interactions, our analysis yields a reasonable collapse with no adjustable parameters.  For long-ranged interactions, there is only one free parameter, $c$, that depends on the arbitrary choice of the original diameter $\sigma$. 

The AWC approximation leading to the collapse shown in fig.~\ref{fig3} also differs from collapses of the dynamics based on the excess entropy.  Based on the ideas of Rosenfeld \cite{rosenfeld} and Gnan \emph{et al.}\cite{gnan}, a number of studies have explored the connection between the dynamics and the excess entropy~\cite{dzugutov,mittal,pond}.   The ratio $T/p_{\textnormal{eff}} \sigma_{\textnormal{eff}}^3$ could perhaps be viewed as a proxy for the excess entropy; for hard spheres the excess entropy depends on $T/p_{\textnormal{eff}} \sigma_{\textnormal{eff}}^3$. At least for some potentials~\cite{pond}, however, what appears to collapse with excess entropy, is the diffusivity made dimensionless by the energy scale of the potential, $\epsilon$.  This is fundamentally different from the relaxation time made dimensionless by the pressure; the difference involves the ratio $p \sigma^3/\epsilon$, which is not fixed. 

In summary, we have extended the results of Xu, et al.~\cite{xu09} by showing that the relaxation times of liquids composed of spheres interacting via finite-ranged repulsions can be mapped up to moderately high pressures onto a system of hard spheres with an appropriate effective diameter: $\tau \sqrt{p \sigma/m} = c f_{\textnormal{HS}}(T/p \sigma^3)$ where $c=1$ for finite-ranged repulsions.  This is accomplished by replacing the soft-sphere system with a system of hard spheres of smaller diameter, $\sigma_{\textnormal{eff}}$, where $\sigma_{\textnormal{eff}}$ is the same value that provides an excellent approximation to structural and other static properties~\cite{andersen71}.   This ``principle of dynamic equivalence" was first noted by Guevara-Rodriguez and Medina-Noyola~\cite{noyola,ramireznoyola} in the context of Brownian particles, and has been generalized in this paper to undamped systems with a wide range of repulsive potentials.

The relaxation time for soft spheres is fairly well described by the hard-sphere curve up to moderate pressures.  This suggests that the  physics that controls the glass transition for hard spheres also controls it for soft-sphere systems.  This is a striking affirmation of the utility of the free-volume concept for soft-sphere liquids.  The free-volume approach to dynamics in liquids, which has been applied to fluids ranging from polymers~\cite{vrentas,victor} to aqueous electrolytes~\cite{frank} to glassforming liquids~\cite{turnbull}, assumes that holes must open up in a dense liquid in order for molecules to rearrange.  In liquids of hard spheres, the concept of a hole is well-defined and the free-volume distribution can be calculated explicitly and efficiently~\cite{sastry98}.  In a system of soft spheres, however, the definition of a hole, and consequently of free volume, has been unclear.  The replacement of soft spheres with hard spheres of smaller diameter, as we have done here, resolves this uncertainty.

Our results suggest that free volume is controlled by two distinct mechanisms in liquids of repulsive spheres.  First, thermal fluctuations perform work against the pressure to open up free volume.  This mechanism is governed by the quantity $T/p\sigma^3$.  Second, at a given $T/p\sigma^3$, the pressure forces soft spheres to overlap so that they behave like hard spheres of smaller diameter.  This mechanism is controlled by $p \sigma^3/\epsilon$.   These two mechanisms combine to determine the ratio of thermal energy over an effective pressure of the system.  The collapse of all of our soft-sphere data onto a hard-sphere relaxation time curve implies that the relaxation time of the soft-sphere liquid depends largely on this ratio.  The next major challenge is to test whether systems with attractive as well as repulsive interactions can be described by generalizations of the two mechanisms for opening up free volume identified here.  Furthermore it would be interesting to see whether our method also works in systems with shear, where temperature over pressure has proven to be a useful control parameter, especially in the limit of small pressures \cite{haxton}.

\acknowledgments
We thank L. Berthier, O. Kogan, T. Schr\o der, T. Witten, N. Xu, and F. Zamponi for helpful discussions. This work was supported by the German Academic Exchange Service (DAAD) within the postdoc program (MS), DE-FG02-05ER46199 (AL and TH), DE-FG02-03ER46088 (SN), NSF-MRSEC DMR-0820054 (SN) and MRSEC DMR-0520020 (TH).

\end{document}